%% file: cp_life_may_20.tex
\documentclass{elsart}
\usepackage[dvips]{graphicx}

\begin{document}
\begin{frontmatter}
\title{A measurement of lifetime differences in the neutral $D$-meson system}
\input{new_author.tex}
\nobreak
\begin{abstract}
Using a high statistics sample of photoproduced charm particles from
the \nobreak FOCUS experiment at Fermilab, we compare the lifetimes of neutral
$D$ mesons decaying via $D^0 \rightarrow K^- \pi^+~{\rm and} ~K^- K^+$
to measure the lifetime differences between CP even and CP odd final
states. These
measurements bear on the phenomenology of $D^0 - \bar D^0$ mixing.
If the $D^0 \rightarrow K^-\pi^+$ is an equal
mixture of CP even and CP odd eigenstates, we measure $y_{\rm CP} = ( \Gamma
({\rm CP~even}) - \Gamma ({\rm CP~odd}))/ ( \Gamma ({\rm CP~even}) +
\Gamma ({\rm CP~odd})) = 0.0342 \pm 0.0139 \pm 0.0074$.  
\end{abstract}
\end{frontmatter}
\newpage
\newpage

This paper contains a comparison of the lifetime of a CP even final
state, $D^0 \rightarrow K^- K^+$, to the lifetime of a CP mixed decay,
$D^0 \rightarrow K^- \pi^+$. The lifetime measurements are made using
high signal-to-background $D^0$ samples consisting of 10\,331 decays
into $K^- K^+$, and 119\,738 decays into $K^- \pi^+$. Throughout this
paper, unless stated explicitly, the charge conjugate is implied when
a decay mode of a specific charge is stated.

If CP violation in neutral $D$-meson decays is negligible, the even CP
and odd CP combinations of the $D^0$ and $\bar D^0$ are 
mass eigenstates with well defined masses and widths. To the extent that
$D^0 \leftrightarrow \bar D^0$ mixing transitions occur, the widths of
the CP even and odd states may differ. This paper reports a new, 
direct measurement of $y_{\rm CP} = ( \Gamma ({\rm CP~even}) - \Gamma ({\rm
CP~odd}))/ (\Gamma({\rm CP~even}) + \Gamma ({\rm
CP~odd}))$. Throughout this paper, we  refer to the width
asymmetry between neutral $D$ CP even and odd eigenstates as $y_{\rm
CP}$ to differentiate it from a mixing parameter (generally called $y$) 
which is the fractional width asymmetry between true mass
eigenstates and could differ from $y_{\rm CP}$ to the extent that
charm decays violate CP symmetry.

Under the assumption that the decay $D^0 \rightarrow K^- \pi^+$ is an
equal CP even - odd mixture, the width difference asymmetry ($y_{\rm
CP}$) is related to the measured lifetimes via:
$$y_{\rm CP}  = { \Gamma ({\rm CP~even}) - \Gamma ({\rm CP~odd})\over 
\Gamma ({\rm CP~even}) + \Gamma ({\rm CP~odd})} = 
{\tau (D^0 \rightarrow K^- \pi^+) \over \tau (D^0 \rightarrow K^- K^+)} -1 
$$
Because $D^0 \rightarrow K^- \pi^+$ is assumed to be a mixed state, a
sizeable width difference between the CP even and odd lifetimes could, in
principle, create a deviation from a pure exponential time
evolution. Given the present limits on $y$ \cite{e791cplife}, this
deviation is safely ignored given the scale of our statistical
precision. Therefore, we will fit both lifetimes assuming a pure
exponential decay.


The data for this paper were collected in the Wideband photoproduction
experiment FOCUS during the Fermilab 1996--1997 fixed-target
run. FOCUS is a considerably upgraded version of a previous experiment, 
E687 \cite{nim}. In FOCUS, a forward multi-particle spectrometer is used to 
measure the interactions of high energy photons on a segmented BeO
target. We obtained a sample of over 1 million fully reconstructed
charm particles in the three major decay modes: $D^0 \rightarrow K^- \pi^+
,~K^- \pi^+ \pi^- \pi^+$ and $D^+ \rightarrow K^- \pi^+ \pi^+$.  We
briefly discuss those aspects of the detector which are
particularly relevant for this analysis.

The FOCUS detector is a large aperture, fixed-target spectrometer with
excellent vertexing and particle identification. A photon beam is
derived from the bremsstrahlung of secondary electrons and positrons
with an $\approx 300$ GeV endpoint energy produced from the 800
GeV/$c$ Tevatron proton beam. The charged particles which emerge from
the target are tracked by two systems of silicon microvertex
detectors. The upstream system, consisting of 4 planes (two views in 2
stations), is interleaved with the experimental target, while the
other system lies downstream of the target and consists of twelve
planes of microstrips arranged in three views. These detectors provide
high resolution separation of primary (production) and secondary
(decay) vertices with an average proper time resolution of $\approx
30~ {\rm fs}$ for 2-track vertices. The momentum of a charged particle
is determined by measuring its deflections in two analysis magnets of
opposite polarity with five stations of multiwire proportional
chambers. Three multicell threshold \v Cerenkov counters are used to
discriminate between electrons, pions, kaons, and protons.  Our
combination of a high quality, high resolution vertex detector, and an
excellent \v Cerenkov system allows us to obtain clean and copious
charm meson samples without the need of $D^*$ tagging although we
employ both a tagged and untagged sample in this analysis.

Throughout this analysis we have chosen cuts designed to minimize
non-charm backgrounds as well as reflection backgrounds from
misidentified charm decays. To minimize potential systematic error, we
use cuts and selection techniques which create very little bias in the
{\it reduced} proper time.  The reduced proper time is a traditional
lifetime variable used in fixed-target experiments which use the
detachment between the primary and secondary vertex as their principal
tool in reducing non-charm background. The reduced proper time is
defined by $t^\prime = (\ell - N \sigma_\ell)/(\beta \gamma c)$ where
$\ell$ is the distance between the primary and secondary vertex,
$\sigma_\ell$ is the resolution on $\ell$, and $N$ is the minimum
``detachment'' cut required to tag the charm particle through its
lifetime. If absorption and acceptance
corrections are small enough that they can be neglected, and
if $\sigma_\ell$ is independent of $\ell$, one can show that the
$t^\prime$ distribution for decaying charmed particles, in the absence
of mixing effects, will follow an exponential distribution. These
assumptions are very nearly true in FOCUS.

\begin{figure}[h!]
\begin{center}
	\includegraphics[height=3.0in]{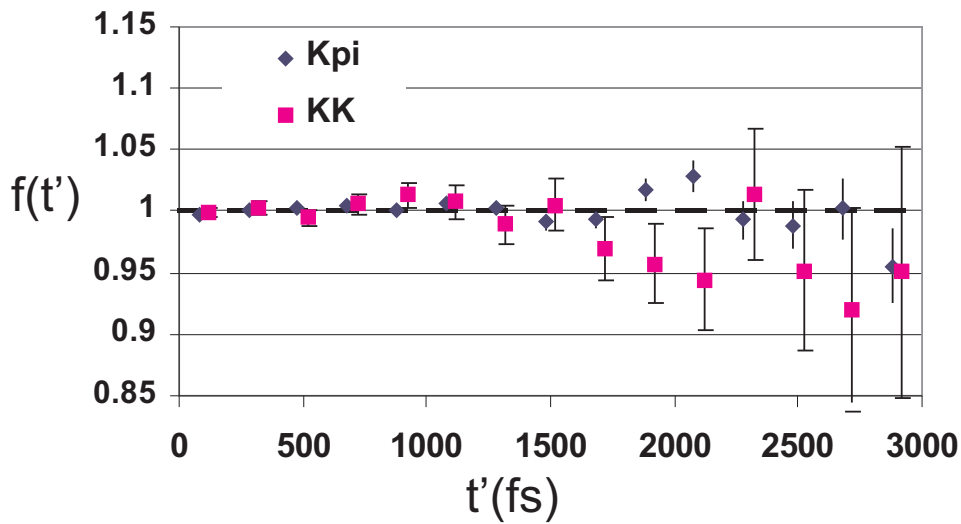}
\end{center}	
\caption{ Monte Carlo correction factors for $D^0 \rightarrow K^-
\pi^+ ~{\rm and}~ K^- K^+$ for $\ell/\sigma > 5$ and $W_\pi -
W_K > 4$. We have offset the $K^- K^+$ points slightly for clarity and
have given them ``flats'' on their error bars.  Monte Carlo corrections
are rather slight with these cuts and the corrections for $D^0
\rightarrow K^- \pi^+$ are the same within errors as those for $D^0
\rightarrow K^- K^+$.  }
\label{ft}
\end{figure}

With a few important differences, many of the basic cuts and analysis
algorithms are described in reference \cite{e687life}. We will
summarize all important analysis issues here as well. Both states
($D^0 \rightarrow K^- \pi^+$ and $K^- K^+$) were obtained using a data
set which required a minimum detachment of the secondary vertex from
the primary vertex of 2.5 $\sigma_\ell$ and a high quality secondary
vertex with a confidence level exceeding 1\%. The primary vertex was
found using a candidate driven vertex finder where a primary vertex
was found by intersecting (nucleating) tracks about a ``seed track''
constructed using the secondary vertex and the reconstructed D
momentum vector. The candidate driven algorithm finds the primary
vertex with relatively high efficiency even at very low detachment. As
we will demonstrate later in Figure \ref{ft}, the use of candidate
driven vertex finder meant that there was essentially no time
dependent efficiency correction required to fit the reduced proper
time to an exponential form. As a result, systematic uncertainties are
greatly reduced since reliance on the Monte Carlo used to compute
efficiency corrections is very minimal.
 
We begin by describing additional vertexing and kinematic cuts
that were used to reduce the background to $D^0 \rightarrow K^- \pi^+
~{\rm and}~K^- K^+$.  These cuts, along with more stringent detachment
cuts and particle identification cuts, were used to obtain our final
sample. To maximize our yields while maintaining good signal-to-noise,
we allow an event to enter our sample through either a {\it
$D^*$tagged} path or an {\it inclusive} path with additional clean-up
cuts. Both paths required that the primary vertex fell within the
boundaries of our segmented target.

The {\it tagged} path admitted any candidate consistent with the decay
$D^{*+} \rightarrow D^0 \pi^+$ by virtue of having a $D^* - D^0$ mass
difference within 3 MeV/$c^2$ of nominal. The very powerful tagging
cut reduced the background to the extent that no additional kinematic
cuts were required. The {\it inclusive} path required that anticipated
proper time resolution for a given event satisfied $\sigma_\ell/(\beta
\gamma c) <60 ~{\rm fs}$, and required that the two tracks did not
have grossly asymmetric momenta ($|P_1 - P_2| / (P_1 + P_2) < 0.70$).
We found from Monte Carlo studies, that background from partially
reconstructed charm decays were often very highly asymmetric and
effectively eliminated by the momentum asymmetry cut.  We found that
the proper time resolution cut was extremely effective at eliminating
long lived backgrounds.  Rather than separately fitting samples from
the two paths, we combined the candidates admitted through either path
into a single sample (in a way that insured no double counting) for
the purpose of fitting.  In doing this, we are implicitly assuming
that the $D^0 \rightarrow K^- K^+$ and $\overline{D^0} \rightarrow K^-
K^+$ have equal lifetimes and hence there is no need to discriminate
between these two decays.

Because misidentified Cabibbo-allowed
decays can be a significant background to the suppressed process $D^0
\rightarrow K^- K^+$, we have studied the charm particle lifetimes
using a variety of \v Cerenkov cuts.  The \v Cerenkov particle
identification cuts used in FOCUS are based on likelihood ratios
between the various stable particle identification hypotheses. These
likelihoods are computed for a given track from the observed firing
response (on or off) of all cells within the track's ($\beta = 1$) \v
Cerenkov cone for each of our three, multicell threshold \v Cerenkov
counters with pion
thresholds of 4.5, 8.4, and 17.4 GeV/c. The probability that a given
track will fire a given cell is computed using Poisson statistics
based on the predicted number of photoelectrons striking the cell's
phototube under each particle identification hypothesis and an
intensity dependent accidental firing rate determined for each of the
300 cells.
The product of all firing probabilities for all cells within the three
\v Cerenkov cones produces a $\chi^2$-like variable called $W_i \equiv
-2 \log(\rm{likelihood})$ where $i$ ranges over the electron, pion,
kaon, and proton hypotheses.

An important \v Cerenkov cut, which we will call {\it
kaonicity}, is defined as $\Delta W_K = W_\pi - W_K$. A
kaonicity cut of $\Delta W_K > 4$
implies that the track we are assigning to the kaon has an
observed \v Cerenkov pattern under the kaon hypothesis that is favored
over that of the pion hypothesis by a factor of $e^2$ = 7.39. Over the
momentum spectrum of typical tracks in FOCUS, pions have significantly
different \v Cerenkov response than kaons, and only a small fraction
(typically $<15\%$) have $\Delta W_K > 0$. Thus, even a mild cut on
the likelihood ratio favoring the kaon hypothesis on the kaon
candidate track reduces backgrounds to Cabibbo-favored decays such as
$D^0 \rightarrow K^- \pi^+ ,~K^- \pi^+ \pi^- \pi^+$ and $D^+
\rightarrow K^- \pi^+ \pi^+$ by a factor of $\approx 10$.

As we will describe later, in order to assess systematic uncertainty,
we varied the kaonicity and detachment cuts used to obtain the final
sample prior to lifetime fitting.  For each fit, a single detachment
and kaonicity cut is used for both the $K^-\pi^+$ and $K^- K^+$ signal
obtained through either the {\it tagged} or {\it inclusive} path. To
further reduce backgrounds for the {\it inclusive} path of the signal,
we required for tracks reconstructed as kaons, that the proton light
pattern is not favored over the kaon hypothesis by more than $W_K - W_P
= 3$. The {\it inclusive} path also required a pion consistency cut
for the ``pion'' track in $K^- \pi^+$ decay that demanded that no
particle hypothesis is favored over the pion hypothesis with a $\Delta
W$ exceeding 2. No additional Cerenkov cuts beyond kaonicity were used
for the {\it tagged} path.

\begin{figure}[h!]
	\includegraphics[height=2.in]{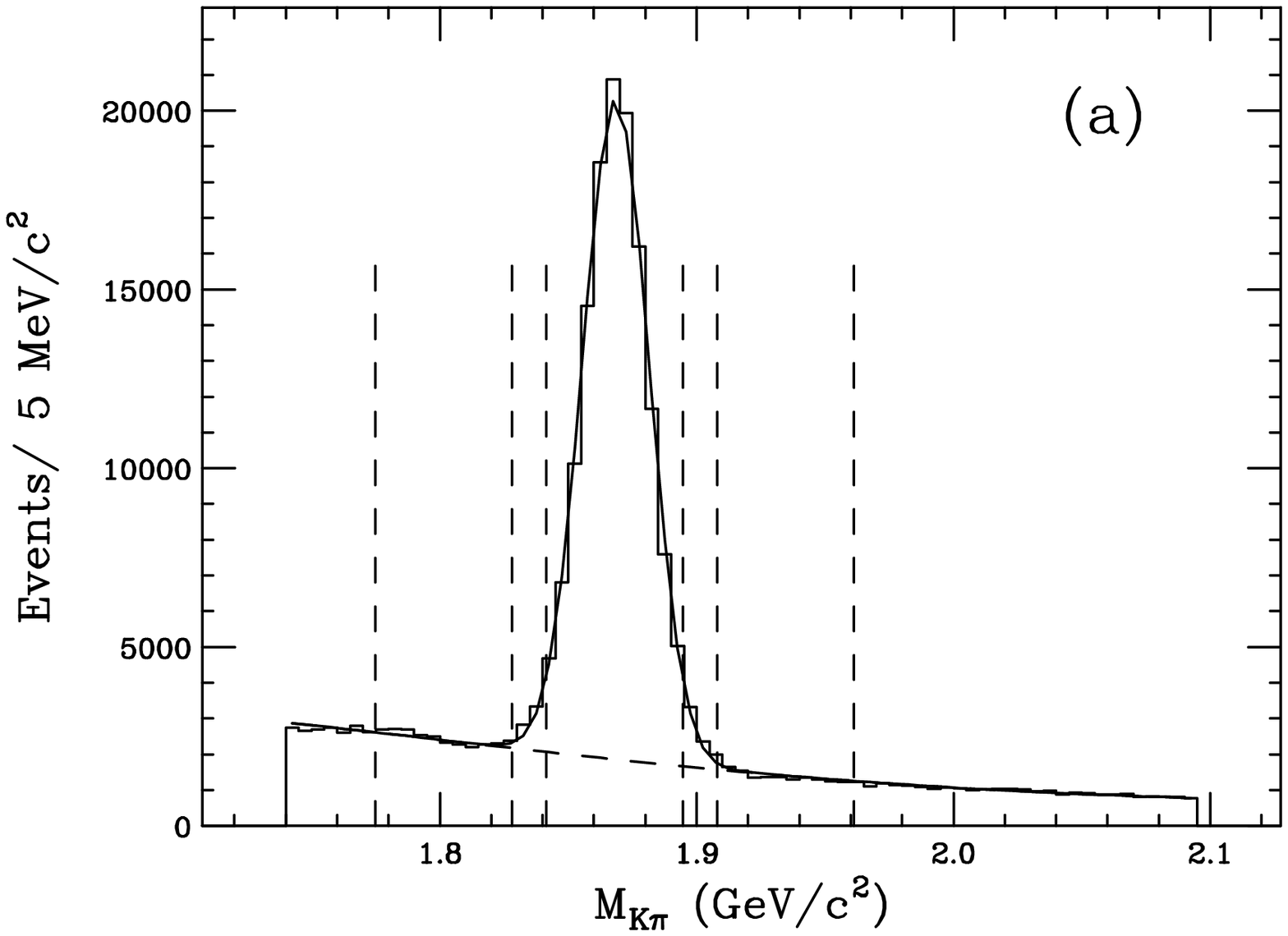}
	\includegraphics[height=2.in]{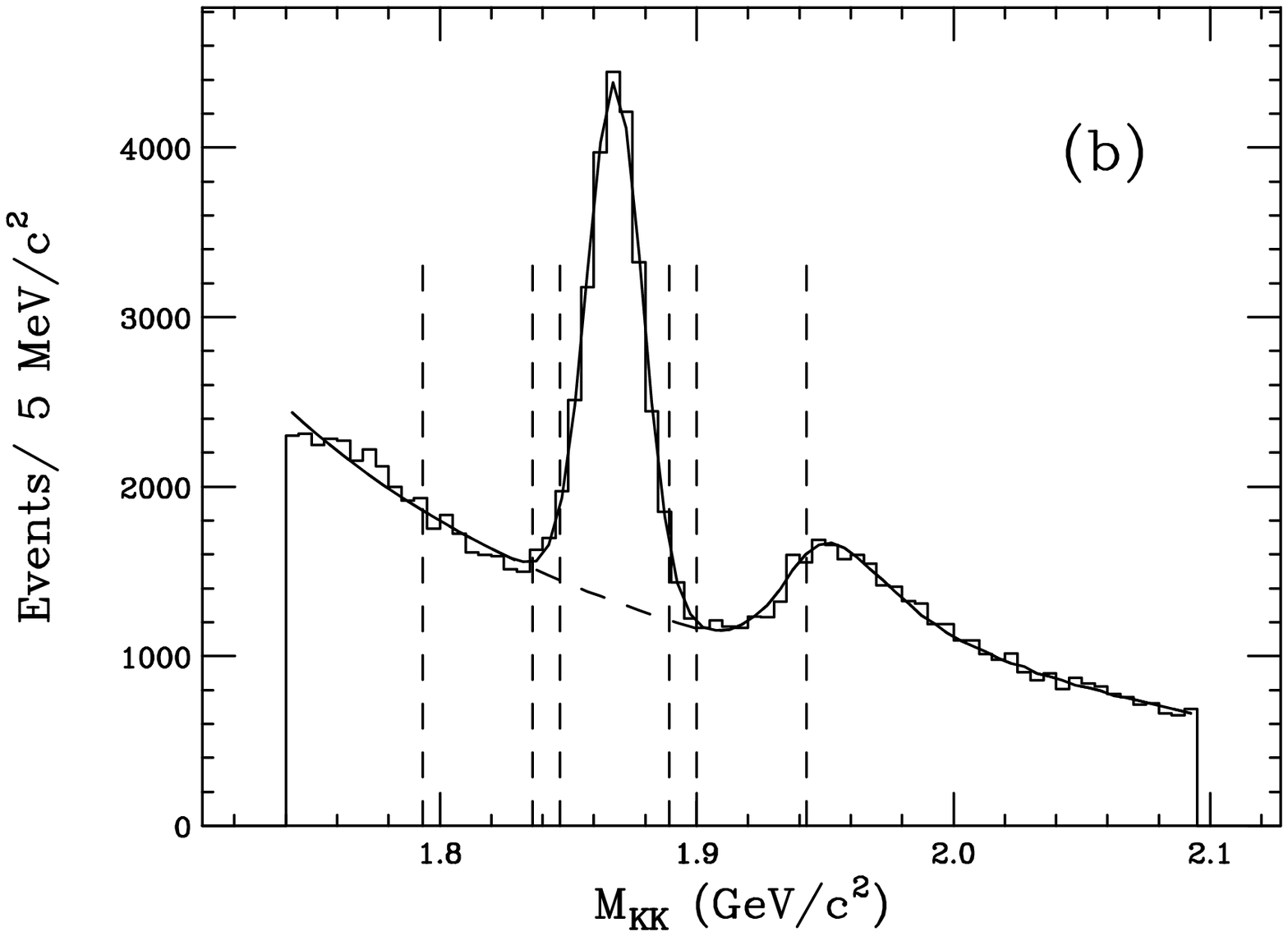}
	\includegraphics[height=2.in]{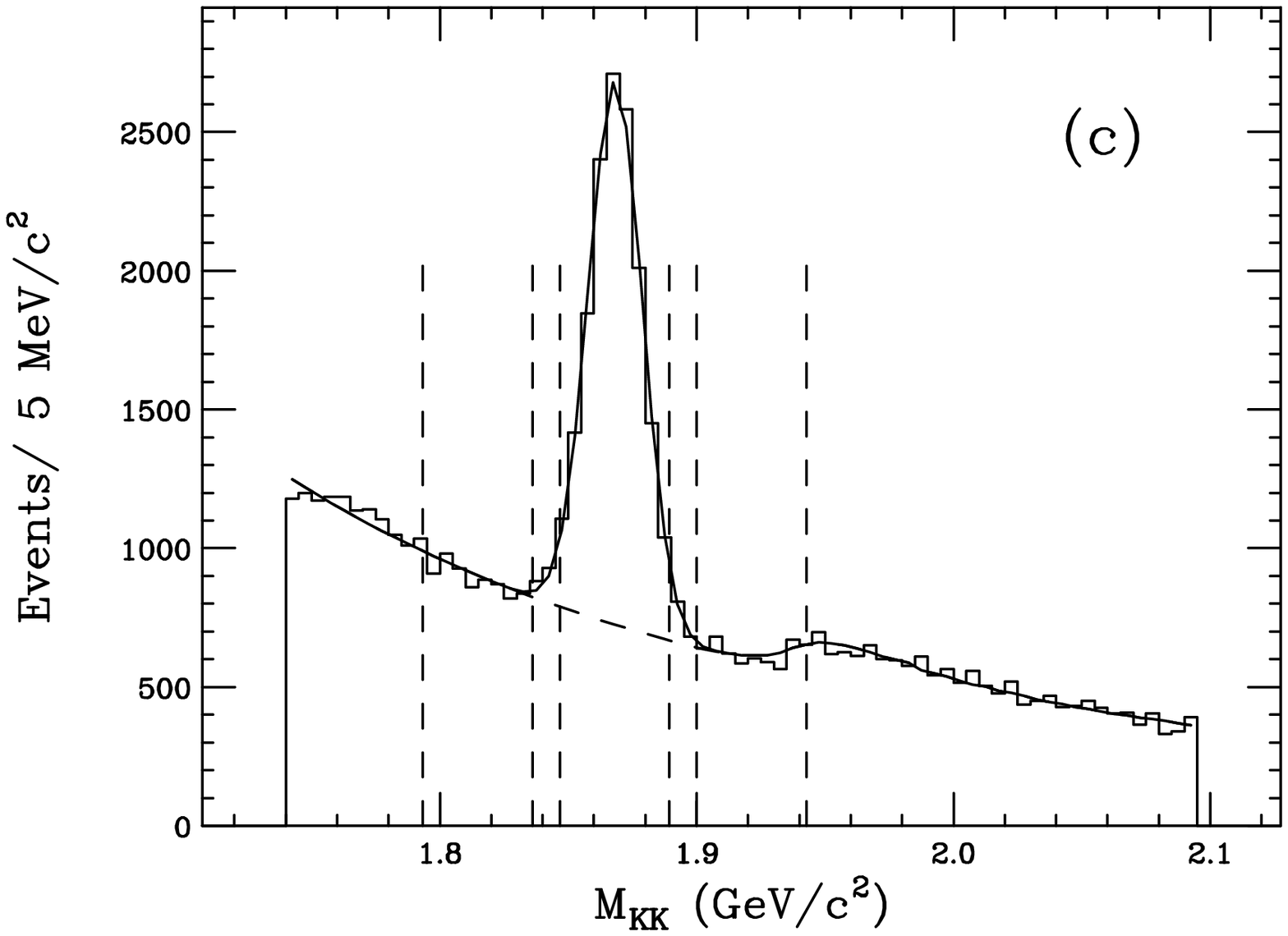}
\caption{ (a) Signal for $D^0 \rightarrow K^- \pi^+$ with a detachment
cut of $\ell/\sigma > 5$ and $W_\pi - W_K > 4$. The yield is 119\,738
$K^- \pi^+$ signal events.\newline Signals for $D^0 \rightarrow K^-
K^+$ with a detachment cut of $\ell/\sigma > 5$. The reflection in the
background at higher masses is due to contamination from misidentified
$D^0 \rightarrow K^- \pi^+$.  (b) Requiring $W_\pi - W_K > 1$, we
obtain a yield of 16\,532 $K^- K^+$ signal events.  (c) Requiring $W_\pi
- W_K > 4$, we obtain a yield of 10\,331 $K^- K^+$ signal events.  The
vertical dashed lines indicate the signal and sideband regions used
for the lifetime and $y_{\rm CP}$ fits.}
\label{signals}
\end{figure}

The mass distributions of Figure \ref{signals} illustrate several of
the signals used in this analysis and specifically the use of these
likelihood-based \v Cerenkov cuts. Figure \ref{signals}(a) shows the
$D^0 \rightarrow K^- \pi^+$ signal obtained after requiring a
kaonicity cut of $\Delta W_K > 4$ for the $K^-$.  Figures
\ref{signals}(b) and \ref{signals}(c) illustrate the use of tight
kaonicity cuts to reduce the $D^0 \rightarrow K^- \pi^+$ reflection
background to $D^0 \rightarrow K^- K^+$. The reduction of the $K^-
\pi^+$ reflection is evident as $\Delta W_K$, applied to both tracks,
is raised from 1 to 4. The $D^0 \rightarrow K^- K^+$ signal yield is
estimated using a Gaussian signal peak over a background consisting of
a 5th-order polynomial to represent general backgrounds along with a
$K^- \pi^+$ reflection line shape taken directly from Monte Carlo but
scaled by a fit parameter to best match the data.

Because the $D^0 \rightarrow K^- K^+$ signals have significant
reflection backgrounds due to misidentified $D^0 \rightarrow K^-
\pi^+$ decays, we use a modified version of the mass sideband
subtraction fitting technique used in our preceding experiment
\cite{e687life}. We will discuss the complete technique here.  We fit the
reduced proper time histogram for $D^0 \rightarrow K^- K^+~{\rm or}~
D^0 \rightarrow K^- \pi^+$ signal region events to a corrected
exponential distribution for the signal added to a background reduced
proper time histogram obtained directly from combinations falling in
either a high or low mass sideband. Because this technique assumes
that the events in symmetrically placed mass sidebands have the same
lifetime evolution as events in the background within the signal
region, it must be modified in light of the $K^- \pi^+$
misidentification reflection shown in Figure \ref{signals} that only
populates the upper sideband. In order to accommodate this reflection,
we subtract the expected contribution from the $D^0 \rightarrow K^-
\pi^+$ reflection from the raw upper sideband, reduced proper time
histogram.  The overall normalization of the reflection contribution
is computed from the integral over the sideband domain of the reflection
peak found through the fit to the $D^0 \rightarrow K^- K^+$ mass
spectrum illustrated in Figure \ref{signals}. The reduced proper time
histogram shape is computed using the lifetime of the $D^0 \rightarrow ~K^-
\pi^+$. This lifetime is taken from a joint fit of the $D^0
\rightarrow K^- K^+ ~{\rm and}~K^- \pi^+$ reduced proper time
distributions where the four fit parameters are (1) lifetime of the
$D^0 \rightarrow ~K^- \pi^+$, (2) the $y_{\rm CP}$ parameter which
relates the $D^0 \rightarrow K^- K^+$ lifetime to the $D^0 \rightarrow
~K^- \pi^+$ lifetime, (3) the background level under the $K^- \pi^+$
signal, and (4) the background level under the $K^- K^+$ signal. The
normalization of the exponential time evolution of the $D^0
\rightarrow K^- K^+ ~{\rm and}~K^- \pi^+$ signals is given by the
total number of events in the signal region minus the background level
fit parameters.

The reduced proper time contributions for the $D^0 \rightarrow ~K^-
\pi^+$,~$D^0 \rightarrow ~K^- K^+$ and misidentified $D^0 \rightarrow
~K^- \pi^+$ reflection are of the form $f(t^\prime) \exp(-t^\prime/
\tau)$ where $f(t^\prime)$ represents efficiency and absorption
corrections to a pure exponential decay with lifetime
$\tau$. The use of a multiplicative ``efficiency''
correction, rather than an integral over a resolution function is
justified since our reduced proper time resolution is less than 1/10
th of the $D^0$ lifetime. Because of the somewhat large (200 fs) bin
widths, we actually integrate the exponential over the domain of the
bin in computing the signal contribution rather than just evaluating
the exponential at the bin center.

A separate $f(t^\prime)$ correction factor, determined using a Monte
Carlo simulation, is used for each of the three exponential
contributions.  Our Monte Carlo simulation includes the Pythia model
for photon-gluon fusion and incorporates a complete simulation at
digitization level of all detector and trigger systems, with all known
multiple scattering and particle absorption effects. We have confirmed
that it accurately reproduces the momentum and $P_\perp$ distribution
for $D$ mesons and the multiplicity and profile of the primary
vertex. The Monte Carlo was run with $20 \times$ the statistics of the
experiment.

Figure \ref{ft} shows the efficiency and absorption corrections
($f(t^\prime)$) obtained using this Monte Carlo for both decay modes
in 200 fs bins of reduced proper time.
The $f(t^\prime)$ function is obtained by dividing the simulated
reconstructed charm yield in each reduced proper time bin by the input
decay exponential integrated over the bin. Figure \ref{ft} shows that
the Monte Carlo corrections are typically less than 5\% for both decay
modes and the corrections for $K^- K^+$ are statistically consistent
with those for $K^- \pi^+$.  Sources of potential $f(t^\prime)$
variation include a minor relative depletion at low $t^\prime$ since
the charm secondaries must lie within the fiducial area of the
downstream microstrips, a depletion at large $t^\prime$ for charm
candidates decaying downstream of the first microstrip detector,
and a slight depletion at low $t^\prime$ since the upstream charm
daughters need to travel through more material before exiting a target
segment for the $\approx 30\%$ of events whose secondary vertex lies
in target material. The charm daughter absorption effect is partially
compensated when one considers absorption of the charm particle itself
which tends to favor low $t^\prime$ for those events produced within
the target material. We assume that the charm absorption
cross section is 1/2 of the cross section for neutrons. Uncertainty in
the charm cross section should cancel when the two charm decay lifetimes
are divided to form $y_{\rm CP}$. 

Because FOCUS uses a segmented target consisting of four 6.75 mm thick
BeO sections, each separated by 10 mm, many decays occur in air. The
charm absorption is minimized in this configuration creating only
minor corrections to the fitted lifetimes. Except for the $\approx 24
\%$ difference between the absorption cross section for kaons and
pions in the momentum range relevant to the $D^0$'s reconstructed in
FOCUS, the minute absorption correction will be common to both
decay modes\cite{PDG} and cancel in $y_{\rm CP}$ . To give a feeling for the
smallness of the absorption correction for lifetimes in FOCUS, we
offer the following example. Computing the $f(t^\prime)$ correction
using a Monte Carlo where the absorption cross sections for charm
secondaries have been scaled by a factor of 60\% relative to their
known values, causes the lifetimes for the $K^- K^+$ and $K^- \pi^+$
to decrease by only about 1.2 fs which implies a very small absorption
systematic uncertainty for $y_{\rm CP}$ compared to the size of the
statistical error.

The background levels in the $D^0 \rightarrow K^- \pi^+$ and $D^0
\rightarrow K^- K^+$ signal region are parameters in the lifetime
fit. We have employed two ways of determining these parameters that
are used to normalize the background contribution to the reduced
proper time histogram in the signal region.  The first method
determines the background levels by finding the number which best fits
the time evolution in the signal region. The second method combines
information from the lifetime evolution with additional information
from the fits to the mass distribution such as those shown in Figure
\ref{signals}. We accomplish this by adding additional likelihood
terms which tend to ``tie'' the total background level to the one deduced
from the mass fit. Specifically for the $D^0 \rightarrow K^- \pi^+$ we
add the log likelihood of a Poisson distribution which ties the $D^0
\rightarrow K^- \pi^+$ level to half of the sum of the number of
candidates in the upper and lower sideband. For the $D^0 \rightarrow
K^- K^+$ we add a $\chi^2$-like likelihood penalty term which ties the
background level to the integral of the polynomial used to represent
the background under the $D^0 \rightarrow K^- K^+$ peak in the signal
region. The incorporation of information from the mass fit tends to
reduce errors by 15--20\% compared to the fits where the background
level is determined from the time evolution alone.

\begin{figure}[h!]
\begin{center}
	\includegraphics[height=3.in]{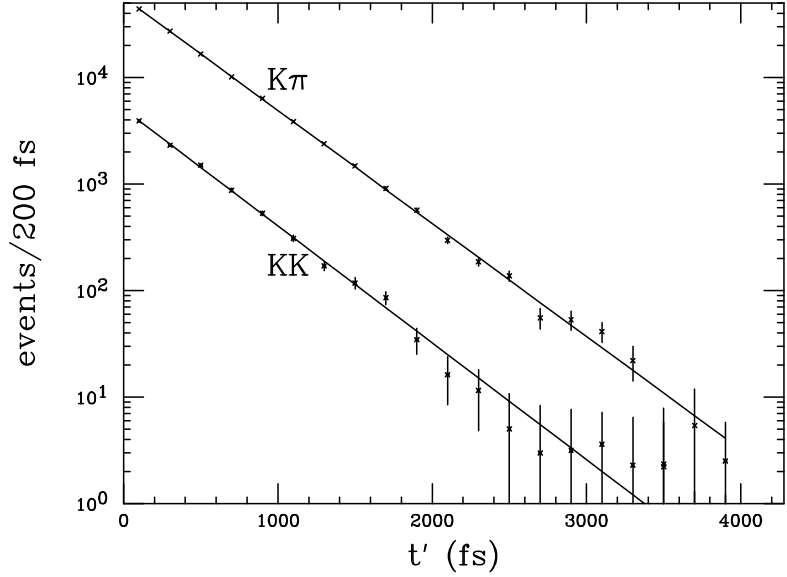}
\end{center}	
\caption{ Signal versus reduced proper time for $D^0 \rightarrow K^-
\pi^+~ ~{\rm and}~~K^- K^+$ requiring $W_\pi - W_K > 4$ and
$\ell/\sigma > 5$. The fit is over 20 bins of 200 fs bin width. The
data is background subtracted and includes the (very small) Monte Carlo
correction.  }
\label{life}
\end{figure}

Figure \ref{life} shows the $t^\prime$ evolution for the $D^0
\rightarrow K^- \pi^+$ and $D^0 \rightarrow K^- K^+$ along with the
predicted number from the lifetime fit. The confidence level for these
fits are 2\% and 55\%, for the $K^- \pi^+$ and $K^- K^+$,
respectively.  From these fits we obtain a lifetime asymmetry of:
\begin{center}
$$y_{\rm CP} = { \Gamma ({\rm CP~even}) - \Gamma ({\rm CP~odd})\over
\Gamma ({\rm CP~even}) + \Gamma ({\rm CP~odd})} = (3.42 \pm 1.39 \pm
0.74) \%$$
\end{center}
and a $D^0 \rightarrow K^- \pi^+$ lifetime of 
\begin{center}
$$\tau (D^0 \rightarrow K^-\pi^+) =  409.2 \pm 1.3 ~{\rm fs} ~{\rm (statistical~error~only)}.$$
\end{center}
Using our value for the fitted lifetime asymmetry, we compute $\tau (D^0 \rightarrow K^-K^+) = 395.7 \pm 5.5 ~{\rm fs} ~{\rm (statistical~error~only)}$.
\begin{figure}[h!]
\begin{center}
	\includegraphics[height=2.5in]{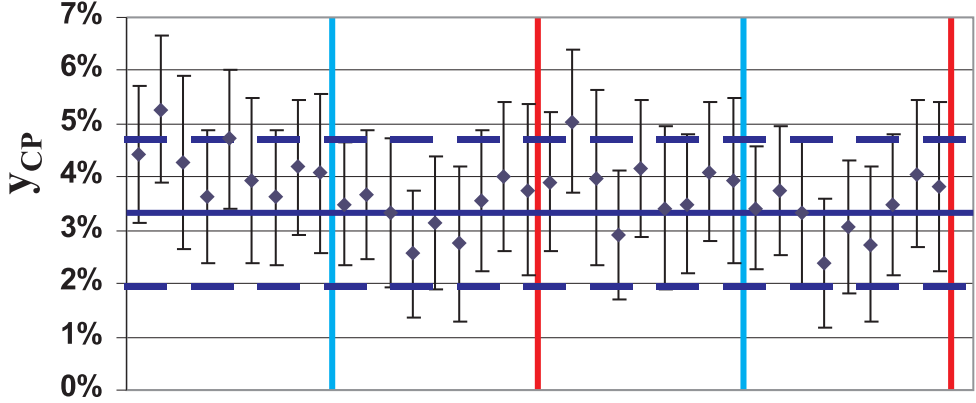}
\end{center}	
\caption{ Stability of the $y_{\rm CP}$ results for 9 sets of clean-up
cuts and 4 different fit options.  The set of 9 cut variants consists
of 3 different kaonicity cuts ($\Delta W_K > 1,2,4$), each with three
different detachment cuts $\ell/\sigma > 5, 7, 9$.  The first 18
values use a 15 bin fit; the last 18 values use a 20 bin fit, where
the bin size remains fixed at 200 fs. The 1st and 3rd set of 9 values
obtain the background level entirely through the time fit. The 2nd and
4th set use the background level which incorporates additional
information from the mass fits shown in Figure \ref{signals}. The RMS
spread in these values is 0.63 \% which is considerably smaller than
our statistical errors. The horizontal lines give our quoted result
on $y_{\rm CP}$ plus or minus its quoted statistical error.}
\label{sys}
\end{figure}
\begin{figure}[h!]
\begin{center}
	\includegraphics[height=2.5in]{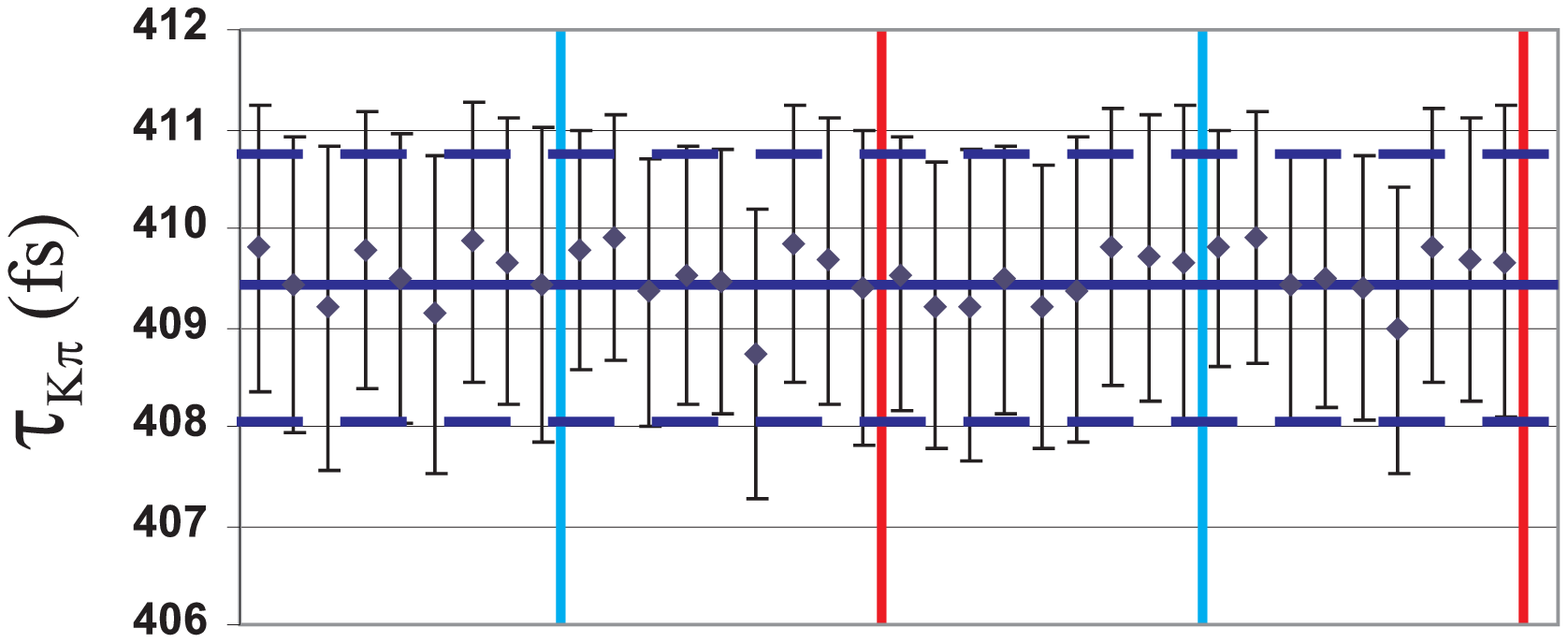}
\end{center}	
\caption{ Stability of the $\tau (K^- \pi^+)$ results for 9 sets of
clean-up cuts and 4 different fit options.  The 36 estimates are
plotted according to the convention of Figure \ref{sys}.  The RMS
spread in these values is 0.28 fs which is considerably smaller than
our statistical errors. The horizontal lines give our quoted result on
$\tau (D^0 \rightarrow K^-\pi^+)$ plus or minus its quoted statistical
error.}
\label{tsys}
\end{figure}

The systematic uncertainty was studied by gauging the
variation of the fitted lifetime estimates as the analysis cuts and
fitting technique are varied. 
Because the Monte Carlo corrections shown in the $f(t^\prime)$ plots
(Figure \ref{ft}) are consistent between the $D^0 \rightarrow K^-
\pi^+$ and $K^- K^+$ sample, one expects the dominant systematic error
on $y_{\rm CP}$ to come from potential differences in the background
under the $K^- K^+$ peak. By varying the minimum $\ell/\sigma$ cut
from 5 to 9 for the case of $K^- \pi^+$ and $K^- K^+$, we
significantly change the relative background level by eliminating
non-charm backgrounds. Defining the signal-to-noise as the
ratio of the signal height to background height at the location of the
Gaussian signal peak, the fits used to measure the lifetimes have
signal-to-noise ratios which range from 8.9 to 19.3 for the $K^-
\pi^+$, from 2.3 to 5.7 for the $K^- K^+$. Changing the \v Cerenkov
log likelihood ratio cuts from $\Delta W_K$ = 1 to 4 significantly
changes the level of charm reflection contamination (as evidenced by
Figure \ref{signals}), and reduces contamination from combinatoric
background. 

Figure \ref{sys} demonstrates the stability of the $y_{\rm CP}$
results for 9 sets of clean-up cuts and 4 different fit options.  The
9 clean-up cut variants considered were 3 different kaonicity cuts
($\Delta W_K > 1,2,4$), each with three different detachment cuts
$\ell/\sigma > 5, 7, 9$.  Each set of 9 points in Figure \ref{sys} are
ordered as ($\Delta W_K > 1$, $\ell/\sigma >5$), ($\Delta W_K > 2$,
$\ell/\sigma >5$), ($\Delta W_K > 4$, $\ell/\sigma >5$) , ($\Delta W_K
> 1$, $\ell/\sigma >7$), ($\Delta W_K > 2$, $\ell/\sigma >7$),
($\Delta W_K > 4$, $\ell/\sigma >7$), ($\Delta W_K > 1$, $\ell/\sigma
>9$), ($\Delta W_K > 2$, $\ell/\sigma >9$), and ($\Delta W_K > 4$,
$\ell/\sigma >9$). The four fit option results, summarized in Figure
\ref{sys}, include varying the lifetime range and the method used to
obtain the background normalization. We show the fitted $D^0
\rightarrow K^- \pi^+$ lifetime for each of the 9 cut variants and 4
fit options in Figure \ref{tsys}.  Our quoted systematic error was
evaluated by first calculating the shifts in $y_{\rm CP}$ for three
different detachment cuts, three different kaonicity cuts, two
different background normalization options, and three different
lifetime fit ranges (10, 15, and 20 bins of 200 fs). These shifts were
then combined in a conservative manner by adding them in quadrature to
obtain the quoted systematic error.

Additional studies, beyond those used to obtain the quoted systematic
error, were made to assess the validity of the Monte Carlo
corrections and the background subtraction technique. We studied the
consistency of the $K^- \pi^+$ lifetime from four samples split
according to momentum and primary vertex location for a variety of
detachment and \v Cerenkov cuts.  These studies include comparing the
$K^- \pi^+$ lifetime for 4 samples split according to the $K^-
\pi^+$ momentum and the location of the primary vertex (in the
upstream two targets versus downstream or upstream two
targets). Although the Monte Carlo ($f(t^\prime)$) corrections for the
high momentum, downstream sample were much severe than those for the
other three samples , the $K^- \pi^+$ lifetimes for all 4 samples were
consistent within errors.

To assess the validity of the background subtraction method, we fit
the $y_{\rm CP}$ values using two versions of the same set of 36 fit
variants summarized in Figure \ref{sys}. The first version used
sideband regions that were set to 1/2 of the width of those shown in
Figure \ref{signals}. The second version restricted events to the
{\it tagged} path only (but with full sideband width illustrated by
Figure \ref{signals}). The signal to background in {\it tagged}-only
sample was 4 times larger than that for our standard {\it tagged} and
{\it inclusive} combined sample. Besides reducing the overall
background level, the tagging requirement should have substantially
reduced any potential backgrounds from partially reconstructed and \v
Cerenkov mis-identified $D^+$ or $\Lambda_c^+$ decays that have a
substantially different lifetime than the $D^o$. Significant
variations in the result when the sideband width is changed might
signify the presence of possible backgrounds (such as those from
possible charm reflections) which have a mass spectrum which is
sufficiently non-linear that the average of the high and low sideband
reduced proper time evolution is no longer an inadequate
representation for the $t^\prime$ evolution of the background in the
signal region.

No problems with the background technique were uncovered by
either of these alternative versions of the 36 standard fits. In all
cases the $y_{\rm CP}$ value was consistent with our standard
value. 80\% of the 36 ``tagged only'' fits and 94\% of the 1/2
sideband fits returned $y_{\rm CP}$ values within the dashed
horizontal lines of Figure \ref{sys}. The remaining fits were at most
3/4 of their error bar away from these lines.


We have presented a new measurement of the lifetime ratio between a CP
even final state, $D^0 \rightarrow K^- K^+$ and a CP mixed decay, $D^0
\rightarrow K^- \pi^+$ of $y_{\rm CP} = (3.42 \pm 1.39 \pm 0.74)\%$. Our
analysis techniques have been designed to minimize the relative
systematic errors between these samples, rather than to obtain the
best statistical error on the $D^0$ lifetime under the assumption of a
pure exponential decay. For example, inclusion of the $D^0 \rightarrow
K^- \pi^+ \pi^- \pi^+$ decay mode would essentially double our
statistics for the $D^0$ lifetime. In addition, there are systematic
error sources such as the overall distance scale error which affects
our absolute lifetime but not the lifetime ratio which is the
principle result reported here.

E791 \cite{e791cplife} measures $\Delta \Gamma = 2(\Gamma_{KK} -
\Gamma_{K\pi})= 0.04 \pm 0.14 \pm 0.05~{\rm ps}^{-1}$. Combining this
and their measurements of the $KK$ and $K\pi$ lifetimes, we obtain a
value of $y_{\rm CP} = (0.8 \pm 2.9 \pm 1.0)\%$ which is consistent
with our measured value of $y_{\rm CP} = (3.42 \pm 1.39 \pm 0.74)\%$.

A more recent result exists from the CLEO Collaboration. CLEO searches
for mixing effects by studying the possible interference of mixing
with direct doubly-Cabibbo-suppressed decays in the time evolution of
$D^{*+} \rightarrow \pi^+ (K^+ \pi^-)$ decays \cite{cleo}. They
report a 95\% confidence level range on a variable they call $y'$ of
$-5.8 \% < y' < 1\%$. If the level of CP violation in charm decays is
negligible, the CLEO $y'$ variable is a rotational transformation of
the $y_{\rm CP}$ variable reported here and a variable which depends
on the CP eigenstate mass difference with the angle of rotation being
due to a strong phase shift.  Theoretical estimates on the size of
this angle differ significantly \cite{phase} making a precise
comparison of our result with the CLEO result impossible at the present time.

Because of our high statistics, the error on $y_{\rm CP}$ reported
here can be reliably interpreted as a Gaussian error for the purposes
of combining with other measurements ({\it e.g.} The $\chi^2$ versus
$y_{\rm CP}$ in a $\pm 1~\sigma$ domain about the fit minimum is well
fit by a parabola).  Our measurement represents the most precise
direct measurement of the neutral $D$ meson CP eigenstate lifetime
difference.  A wide range of Standard Model and non-Standard Model
predictions on mixing through the width difference appear in the
literature \cite{cptheory}.

We wish to acknowledge the assistance of the staffs of Fermi National
Accelerator Laboratory, the INFN of Italy, and the physics departments
of the collaborating institutions. This research was supported in part
by the U.~S.  National Science Foundation, the U.~S. Department of
Energy, the Italian Istituto Nazionale di Fisica Nucleare and
Ministero dell'Universit\`a e della Ricerca Scientifica e Tecnologica,
the Brazilian Conselho Nacional de Desenvolvimento Cient\'{\i}fico e
Tecnol\'ogico, CONACyT-M\'exico, the Korean Ministry of Education, and
the Korean Science and Engineering Foundation.

\end{document}

%% file: new_author.tex
The FOCUS Collaboration

\author[davis]{J.M.~Link},
\author[davis]{V.S.~Paolone\thanksref{atpitt}},
\author[davis]{M.~Reyes\thanksref{atmex}},
\author[davis]{P.M.~Yager},
\author[cbpf]{J.C.~Anjos},
\author[cbpf]{I.~Bediaga},
\author[cbpf]{C.~G\"obel\thanksref{aturug}},
\author[cbpf]{J.~Magnin\thanksref{atbogota}},
\author[cbpf]{J.M.~de~Miranda},
\author[cbpf]{I.M.~Pepe\thanksref{atbahia}},
% ...

\address[davis]{University of California, Davis, CA 95616}
\address[cbpf]{Centro Brasileiro de Pesquisas F\'\i sicas, Rio de Janeiro,
RJ, Brazil}
\address[cine]{CINVESTAV, 07000 M\'exico City, DF, Mexico}
\address[cu]{University of Colorado, Boulder, CO 80309}
\address[fnal]{Fermi National Accelerator Laboratory, Batavia, IL 60510}
\address[fras]{Laboratori  Nazionali di Frascati dell'INFN, Frascati, Italy,
      I-00044}
\address[ui]{University of Illinois, Urbana-Champaign, IL 61801}
\address[iu]{Indiana University, Bloomington, IN 47405}
\address[koru]{Korea University, Seoul, Korea 136-701}
\address[milan]{INFN and University of Milano, Milano, Italy}
\address[anc]{University of North Carolina, Asheville, NC 28804}
\address[pavia]{Dipartimento di Fisica Nucleare e Teorica and INFN,
Pavia, Italy}
\address[pr]{University of Puerto Rico, Mayaguez, PR 00681}
\address[sc]{University of South Carolina, Columbia, SC 29208}
\address[ut]{University of Tennessee, Knoxville, TN 37996}
\address[vandy]{Vanderbilt University, Nashville, TN 37235}
\address[wisc]{University of Wisconsin, Madison, WI 53706}
\address[yon]{Yonsei University, Seoul, Korea 120-749}

\thanks[atpitt]{Present Address: University of Pittsburgh, Pittsburgh,
PA 15260}
\thanks[atmex]{Present Address: Instituto de F\'\i sica y
Matematicas, Universidad Michoacana de
San Nicolas de Hidalgo, Morelia, Mich., Mexico 58040}
\thanks[aturug]{Present Address: Instituto de F\'\i sica, Facultad de
Ingenier\'\i a, Univ. de la Rep\'ublica, Montevideo, Uruguay}
\thanks[atbogota]{Present Address: Universidad de los Andes, Bogota,
Colombia}
\thanks[atbahia]{Present Address: Instituto de F\'\i sica, Universidade
Federal da Bahia, Salvador, Brazil}
\thanks[atmilan]{Present Address: INFN sezione di Milano, Milano,
Italy}
\thanks[atpr]{Present Address: University of Puerto Rico, Mayaguez,
PR  00681}
\thanks[atpub]{Present Address: Instituto de F\'{\i}sica, Universidad
Auton\'oma de Puebla,  Puebla, M\'exico}
\thanks[atncar]{Present Address: National Center for Atmospheric Research,
Boulder, CO} 
\thanks[atlbl]{Present Address: Lawrence Berkeley Lab, Berkeley, CA
94720}
\thanks[atbrescia]{Present Address:
Dipartimento di Chimica e F\'\i sica per l'Ingegneria e per i Materiali,
Universita' di Brescia  and  INFN sezione di Pavia}
\thanks[atromania]{Present Address: Nat. Inst. of Phys. and Nucl. Eng., Bucharest,
Romania}
\thanks[atcu]{Present Address: University of Colorado, Boulder, CO 80309}
\thanks[atlucent]{Present Address: Lucent Technology}
\thanks[atagusta]{Present Address: Augusta Technical Inst., Augusta, GA
30906}
\thanks[atadams]{Present Address: Adams State College, Alamosa, CO 81102} 